\documentclass[aps,prb,twocolumn,groupedaddress,intlimits,amsmath,amssymb,floats]{revtex4}
\usepackage{bm}
\usepackage[final]{graphicx}
\usepackage{lscape}
\usepackage{rotating}
\usepackage{epsfig}
\usepackage{color}
\usepackage{tabularx}
\usepackage{array}
\usepackage{hyperref}

\definecolor{DarkBlue}{rgb}{0.1,0.1,0.8}
\definecolor{Red}{rgb}{0.9,0.0,0.1}
\definecolor{Green}{rgb}{0.0,0.99,0.0}

\begin{document}
\title{Multiplets and Crystal Fields: 
Systematics for X-Ray Spectroscopies}
\date{\today}
\author{F. Vernay and B. Delley}
\affiliation{Condensed Matter Theory Group, Paul Scherrer Institut, 
CH-5232 Villigen PSI, Switzerland
}
\begin{abstract}
An easily accessible method is presented that permits to calculate
spectra involving atomic multiplets
relevant to X-ray Absorption Spectroscopy (XAS) and Resonant 
Inelastic X-ray Scattering (RIXS) experiments.
We present specific examples and compare the calculated spectra with 
available experimental data.  
\end{abstract}
\maketitle

\section{Introduction}
Since the early days of quantum mechanics and 
atomic spectroscopy, it has been realized that electron-electron 
interactions and spin-orbit coupling were at the origin of the splitting 
of the electronic shells into {\it multiplet levels}.\cite{condon} 
As they originate from a many-body problem, these levels are relatively 
straightforward to compute and understand analytically as long as the number 
of involved particles remains small. Relatively rapidly one is forced to require 
computer assistance. To this effect, Cowan\cite{cowan_book} implemented 
a systematic computer code in order to solve multiplet structures 
arising in atomic spectra. 

Core spectroscopies have gained interest in solid state physics 
in the eighties as it has been shown that they can provide deep  
insights about the electronic structure of the materials. As the 
processes involve one or more open-shell they exhibit signatures 
of multiplet structures,\cite{moser,thole+cowan,thole2} these 
structures being sensitive to the presence of localized 
moments and hybridization with the ligands. Later on, using a multiplet-based 
code, a systematic study of XAS in $3d$ transition-metal compounds within 
a cubic crystal field environment as been performed by de Groot 
{\it et al}.,\cite{degroot_xtal} showing good agreement with experiments.  
Indeed, soft x-ray XAS\cite{degroot} and also RIXS\cite{kotani} are 
linked to local physics: the absorption of a photon and creation 
of a localized core-hole opens up a shell and therefore the multiplet 
structure becomes apparent in the spectra. 

With the development of sources and progress in optics,  
X-ray spectroscopies such as XAS and RIXS, especially in the soft X-ray 
regime,\cite{saxs} became tools of choice to investigate 
transition correlated materials such as, for instance, 
cuprates\cite{braico,justina}  
or vanadates:\cite{yvo,thorsten,vanadate} the current resolving power 
allows to investigate multiparticle excitations like it was predicted 
in different theoretical contexts for inelastic X-ray 
scattering.\cite{us,donkov,vdb,forte}
 From here we see that it becomes 
crucial, while interpreting the experimental data, to have a systematic, 
user-friendly and transparent way of computing the multiplet spectra in order 
to disentangle in the experiment the information arising from single-particle 
excitations from the information relevant to collective excitations.

As noticed earlier, atomic multiplets code already exist in the 
X-ray spectroscopy community,\cite{cowan,thole,missing} they are based on Cowan's 
pionneering work and make use of symmetries in a very educated way. However, 
for a general audience and especially for newcomers to the field of 
X-ray spectroscopy, we felt that with nowadays computing power we had 
the opportunity of writing a less expert-oriented code, in that sense that 
the diagonalization of the Hamiltonian can be done brute force and the 
local crystal field can be implemented in a straightforward manner. 
Atomic-multiplet calculations for arbitrary symmetries have already been 
reported,\cite{mirone} however the implimentation of the crystal-field in 
the latter approach still requires the introduction of free parameters.  
The aim of this paper is both to present our approach which combines the 
Density Functional Theory (DFT) -based solution form the radial Dirac 
equation to a full diagonalization procedure in order to compute the 
multiplet structures as well as the associated XAS and RIXS spectra
by treating all interactions (electron-electron, spin-orbit coupling 
and crystal-field) on the same footing.  
 
The paper is organized as follows: in a first part we remind the reader 
about atomic multiplets as described extensively in 
Refs.[\onlinecite{condon,cowan_book,degroot,balhausen}], we discuss how to 
implement the crystal field in our calculation. From these basic steps we  
are in a position to compute explicitely XAS and RIXS spectra~; 
two sections are devoted to these issues where we also discuss the 
implementation of the optical selection rules in the dipolar approximation.   

\section{Atomic Multiplets: overview}
\subsection{Electronic interactions and Hilbert space}
By solving the one-particle Dirac equation we know that the electronic states 
are determined by quantum numbers $n$ and $j$. In a 
non-relativistic treatment we can define an orbital quantum number $l$ and 
{\it shells} for the orbitals (like $s$, $p$, $d$, $f$ for 
respectively $l=0,1,2,3$) such that these 
shells have a $(2l+1)$-fold orbital degeneracy times the Kramers 
degeneracy for the spin degree of freedom. For instance, we see that 
$(2\times 2+1)\times 2=10$ electrons can fit in a $d$-shell. 

If we consider the case of a non-hydrogenic atom, with more than one electron, 
the electronic eigenstates and eigenvalues  are not simply labeled by the 
quantum numbers $n$ and $l$ because the electrons interact electrostatically 
with each-other (and also {\it via} spin-orbit coupling).  
As long as a shell is empty or fully occupied it remains  
symmetric and the orbitals of a given shell remain degenerate. 
However, for an open shell electrostatics and spin-orbit interactions will 
split the shell and form a multiplet structure. 

Hence, the Hamiltonian describing the system of electrons is given by~:
\begin{equation}\label{hamilton}
{\mathcal H}=\sum_{i}\epsilon_i + \sum_{i<j} \frac{e^2}{|{\bf r}_i-{\bf r}_j|}
\end{equation} 
where the first term is a single-particle diagonal operator containing 
the kinetic and automatically the spin-orbit energy since we use  
solutions from the Dirac equation.
The second term is simply the 
sum over all pairs of electrons for electrostatics interactions. 
This last term being off-diagonal we see that we end up with a 
diagonalization problem in order to deduce the multiplet structure.  

In the present paper we are aiming at describing soft X-ray absorption and 
emission processes involving two shells. Hence, we can define the Hilbert 
space associated to this problem by only considering the electrons involved 
in these two shells. 
For a shell determined by the quantum 
numbers ($n,\ l$) and containing $k$ electrons, the dimension of the Hilbert 
space is~:
\begin{equation}  
{\mathcal N}(l,k) = \frac{[2(2l+1)]!}{[2(2l+1)-k]!\ k!}
\end{equation}
For instance 2 electrons in a $p$-shell gives ${\mathcal N}(2,2)=15$ states, 
 2 electrons in a $d$-shell gives ${\mathcal N}(3,2)=45$ states.  
If more than one shell is opened, each shell being independent, the overall 
size is given by the product ${\mathcal N}=\prod_{i}{\mathcal N}_i$. 

While the case of transition metals can be handled on standard desktop 
computers, it is worth mentioning the case of lanthanides. 
For instance, the largest Hilbert space is given for the case 
of Gd $4f^6 5d^1\ \to 4f^5 5d^2$ and has a dimension 
135135 for its final-space, it necessitates a too large amount of memory. 
Fortunately, most of the Hilbert spaces involved for soft X-ray scattering 
experiments are of the order of Sm at M$_{4,5}$-edge\cite{thole2} 
$4d^{10} 4f^6\ \to 4d^9 4f^7$  which are of respective dimensions 3003 and 
34320 ($\sim$ 9GB), such sizes being available on modern computers.

Care has to be taken while generating the Hilbert space~; the space is 
obtained by taking the direct product of the one-particle states in the Fock 
space, yet, electrons are fermions, therefore they obey Pauli exclusion 
principle and they anticommute.  
\subsection{Lattice and Crystal Field}
So far, Eq.(\ref{hamilton}) did involve the matrix elements for a 
multiplet formed by the electron-electron interaction and the spin-orbit 
coupling of a single isolated atom, but we have to keep in mind that 
the aim of this paper is to simulate multiplet structures of ions in solids. 
Therefore, we consider in this section an ion embedded in a cluster; the 
considered ion feels the electrostatic {\it crystal field} potential 
created by the other ions forming the cluster. 

The crystal field splitting in simple geometries (cubic or its subgroups) 
can be deduced from symmetry arguments. In our approach we choose to 
implement the crystal field in the full-diagonalization procedure. 
For the implementation of the crystal field we assume a point charge model~: 
the ion with core excitation is surrounded by point charge ligands. 

The advantage of solving the problem of a cluster containing 
the ion and the neighboring ligands lies in two facts: {\it i)} there is 
no need to start from a cubic symmetry
but the position of each anion can directly be entered in the program~; 
{\it ii)} the crystal-field strength is not a free-parameter but 
the code allows to be predictive.

The potential created by the ligands of charge $q_i$ at a distance $R_i$ on 
the orbital $r$ is given by the electrostatics~:
\begin{equation}
V({\bf r})=\sum_{i=1}^N\frac{q_ie^2}{|{\bf r}-{\bf R}_i|}
\end{equation}
assuming that $R_i>r$ the previous expression can be expanded on the 
spherical harmonics basis, 
as we will see in the appendix (Eq.\ref{electro}). 

The introduction of a crystal field in the problem explicitely breaks the 
spherical symmetry of the problem, therefore we have to choose a local 
cartesian basis $({\bf e_x,e_y,e_z})$.  
A typical distance $d$ (in \AA) is introduced defining the unity of the 
coordinates system $({\bf x,y,z})$, such that the surrounding 
ions are represented by their positions in this basis and the formal charges 
that they carry in units of $e$.

\begin{table}[t]
   \caption{\label{tablo} Positions 
of the ligands for a 4\% uniaxially distorted perovskite. 
The position are in units of the typical distance $d=$1.96\AA, 
the formal charges are in units of $e$.}
\begin{tabular}{c|c|c|c}
   x  & y & z & charge \\
  \hline
  \hline
  -1.00 &  0.00 &  0.00 & -2.00\\
   1.00 &  0.00 &  0.00 & -2.00\\
   0.00 & -1.00 &  0.00 & -2.00\\
   0.00 &  1.00 &  0.00 & -2.00\\
   0.00 &  0.00 & -0.96 & -2.00\\
   0.00 &  0.00 &  0.96 & -2.00\\   
\end{tabular}
\end{table}
\begin{figure}[]
\includegraphics*[width=7cm,angle=0]{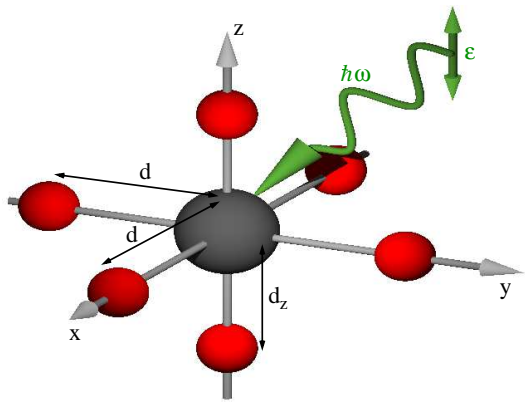}
\caption{(Color online) Ion in a distorted 
octahedral environment: the ion with core excitation is 
in grey while the red spheres represent the ligands, in green 
the incoming photon with its associated polarization vector 
$\bm \epsilon$.}\label{octa}
\end{figure}

A straightforward example is a
$d^1$ ion (without spin-orbit coupling) in the perovskite 
structure. Considering the octahedral environment only, 
for the undistorted case, two levels should be observed while, with 
a slight distortion of the octahedron, the degeneracy of the 
$t_{2g}$ level is partially lifted and the two $e_g$ levels split-up as well. 
The schematic shown in Fig.\ref{octa} defines the $({\bf x,y,z})$ basis 
chosen to define the local crystal field such that in units of $d$, 
the 6 ligands are located at $(\pm 1,0,0)$, $(0,\pm 1,0)$ and 
$(0,0, \pm d_z/d)$. The positions and charges of the ions corresponding 
to this case is summarized in Table~\ref{tablo} and reflects the 
input parameters of the code. 

\begin{figure}[]
\includegraphics*[width=\columnwidth,angle=0]{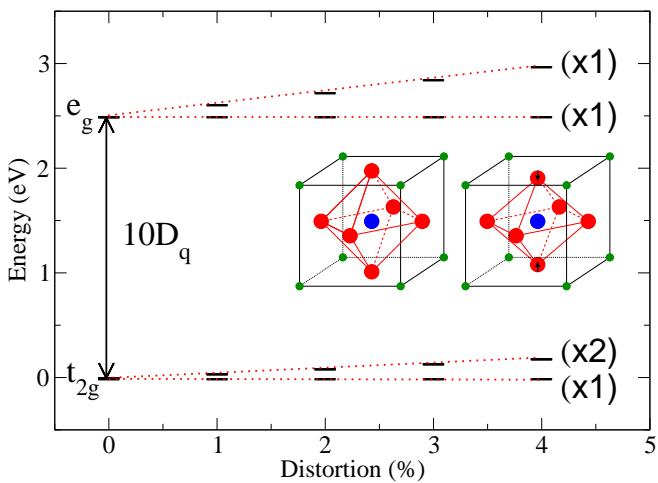}
\caption{(Color online) 
Partial lifting of the degeneracies as an effect of the distortion, 
the orbital degeneracy is indicated in parenthesis. 
{\it Inset:} Undistorted and distorted perovskite structure in blue 
Ti$^{3+}$, in red O$^{2-}$ and in green La$^{3+}$ ions.}\label{pero}
\end{figure}

The case of a distortion in a perovskite structure of the type of 
LaTiO$_3$ is displayed in Fig.~\ref{pero}. 
It is relatively easy to understand qualitatively what is happening in 
Fig.~\ref{pero}: for the cubic symmetry, the splitting is observed because 
the $e_{g}$ orbitals ($d_{x^2-y^2},d_{3z^2-r^2}$) have their lobes pointing 
towards the O$^{2-}$, therefore they are energetically disfavored 
compared to $t_{2g}$ orbitals. When the distortion takes place, the two 
apical oxygens are moved closer to $d_{3z^2-r^2}$ orbital on one hand, and 
to $d_{xz},d_{yz}$ orbitals on the other hand, leading to the splitting 
of both sub-shells.   

It may be of interest to notice that traditionally, the crystal field 
splitting between $t_{2g}$ and $e_g$ levels is referred to as $10D_q$, where 
the parameter $D_q$ is defined by convention~:
\begin{equation}
\begin{array}{rcl}
6D_q &=& \int \psi^\star(e_g) V \psi(e_g)\ d\tau\\
-4D_q &=& \int \psi^\star(t_{2g}) V \psi(t_{2g})\ d\tau
\end{array}
\end{equation}
such that the barycenter of the splitting lies at 0. $D_q$ is thus taken as a
measure of the crystal field strength and often adapted to fit experiments. 

Compared to a summation over the whole lattice, the inclusion of the 
crystal field by only considering the surrounding environment of an ion 
remains relevant for the present investigation. Indeed, we can include 
many ions --and not only the closest ones-- such that the symmetry of the 
crystal field is preserved. Once the symmetry independent parameters of 
the crystal field are exhausted, addition of further more neighbor shells 
has a minor quantitative effect. Screening of the bare crystal field 
may bee taken into account by a scaling factor. 

\subsection{Effect of Ionization}\label{ionization}
\begin{figure}[t]
\includegraphics*[width=\columnwidth,angle=0]{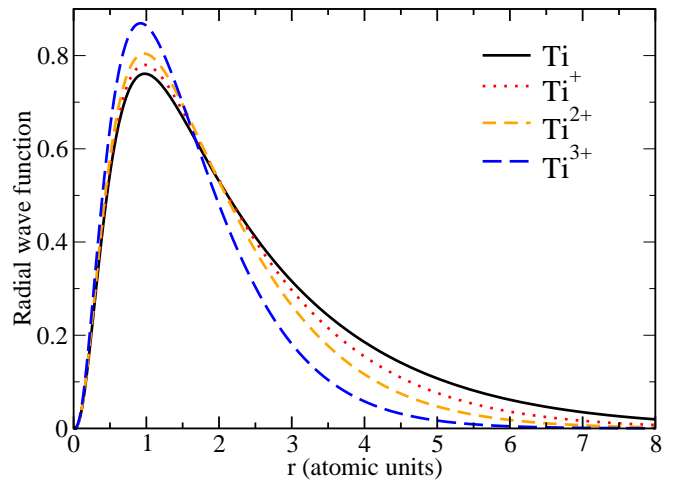}
\caption{(Color online) 
Plot of the radial part of the $d$-wave function for Ti as 
a function of $r$ in atomic units~: the tail of the wave function shrinks 
with the ionization.}\label{radial}
\end{figure}

The ionization has a major impact on the radial part of the 
wave-functions. As a consequence of taking one or many electrons out 
of an atom, the remaining electrons feel a less screened nuclear potential 
and the wave-functions contract around the nucleus, this effect can easily 
been seen in Fig.~\ref{radial}.    

To be quantitative, for this specific case, the expectation value 
$\langle r \rangle$, has a variation of about $30\%$ between the neutral 
titanium ion and Ti$^{3+}$ ion. At first glance, this does not seem 
to be dramatic but this quantity enters in powers of $l$ in the 
evaluation of the crystal field, and we see that taking formal charges too 
seriously can have a strong impact leading to underestimating the crystal 
field strength. To avoid this situation we consider systematically neutral 
ions while computing the radial wave-functions and allow for a scaling 
of the crystal field strength if necessary. Indeed, self-consistent 
band-structure calculations suggest that neutral atoms represent the charge 
density around an atom more realistically than ionic. We checked on different 
transition metal oxides (at the L-edge) that this approximation predicts   
reasonably well crystal-field strengths and does not necessitate the 
introduction of other fitting parameters. 
\subsection{Spin-orbit coupling}
Previous atomic multiplet calculations, even recent ones,\cite{mirone} 
often considered solutions of the nonrelativistic Schr\"odinger equation, 
meaning that the spin-orbit coupling had to be introduced by hand in the 
Hamiltonian matrix. Since we are willing to treat all interactions on the 
same footing, the most  straightforward approach is to consider the 
solutions of the Dirac equation. Hence,  the spin-orbit coupling manifests 
itself in two ways: 
in Eq.(\ref{hamilton}) the values $\epsilon_i$ depend on the total angular 
momentum $j$ as well as the radial wave-functions do. 
From our experience and as we will show in the next section, for 
X-ray absorption spectroscopies, this treatment of the spin-orbit 
coupling allows a relatively good first-principle prediction of the 
L$_2$-L$_3$ edge splitting. Yet, we allow for an arbitrary scaling of this 
coupling which enables the user to have a better match with experimental data 
if needed.  
\subsection{Polarization}
As noticed earlier, the introduction of the crystal field breaks the 
spherical symmetry of the pure atomic case. The description of 
X-ray scattering experiments necessitate to discuss in this subsection 
the implications of the crystal field on polarized photon experiments. 
The incoming photons are defined by both their energy $(\hbar\omega)$ 
and polarization $\bm \epsilon$, 
the polarization property plays a central role in the scattering process: 
depending on the polarization a transition is dipole-allowed or 
dipole-forbidden as we will show in the next section. Thus, it 
is crucial to implement the polarization in the problem. The most natural way 
of introducing the polarization is by referring to its projection 
$({\bm \epsilon_x,\bm \epsilon_y,\bm \epsilon_z })$ on the same local basis 
$({\bf x,y,z})$ which has been defined for the position of the
surrounding ions. For the specific example of Fig.\ref{octa}, 
the polarization associated to the incoming photon is 
${\bm \epsilon}=(0,0,1)$.  

The particular case of unpolarized photons, or experiments done on powder 
samples, we can always recover this limit by doing an incoherent superposition 
of the different polarizations. 
\section{X-ray Absorption Spectroscopy (XAS):}
\subsection{Description}
X-ray absorption spectra are obtained by shining an X-ray photon on a 
material, the photon is absorbed and gives rise to a transition of a 
core-electron into an excited level. 
Thus a {\it core-hole} is formed and therefore at least 
one shell is opened resulting the formation and investigation of 
the multiplet structure.

The absorption spectra can be evaluated with the Fermi 
Golden rule~:
\begin{equation}\label{abs1}
I(\omega)=\sum_i|\langle\psi_i|\hat{{\mathcal O}}|\psi_0\rangle|^2
\delta(\omega+E_0-E_i)\ ;
\end{equation}
where $(|\psi_0\rangle,\ E_0)$ refers to the ground-state and its eigenvalue, 
$(|\psi_i\rangle,\ E_i)$ to the final state and final energy, whereas 
$\omega$ is the energy of the absorbed photon. The operator 
$\hat{\mathcal O}$ represents the transition operator which is treated in 
the dipolar approximation. 

Equation (\ref{abs1}) has its limitations both from an 
intrinsic and experimental point of view : indeed the core-hole has a 
finite intrinsic lifetime and the experiment has of course a finite
resolution. These two facts lead to spectral broadening which we can 
mimic by a Lorentzian and Eq.(\ref{abs1}) becomes~:
\begin{equation}
I(\omega)=\sum_i|\langle\psi_i|\hat{{\mathcal O}}|\psi_0\rangle|^2
\frac{\Gamma/\pi}{(\omega+E_0-E_i)^2+\Gamma^2}
\end{equation}  
In practice, the broadening is partly due to the experimental resolution 
but it should be mentioned that  
the broadening term $\Gamma$ may not remain a constant but can be a function 
of $\omega$ and $\psi_i$~: this reflects hybridizations and vibrational 
effects which depend on the orbitals involved in the states as well as 
electron-electron scattering. 

To keep a straightforward approach to our calculation we will assume a 
finite but constant broadening which can be tuned to match the experimental 
limitations. The most striking consequence of this assumption lies in the 
fact that the relative intensities between the peaks in our simulations may 
be slightly different from what would experimentally be observed.  
\subsection{Dipolar Approximation}
\begin{figure}[t]
\includegraphics*[width=\columnwidth,angle=0]{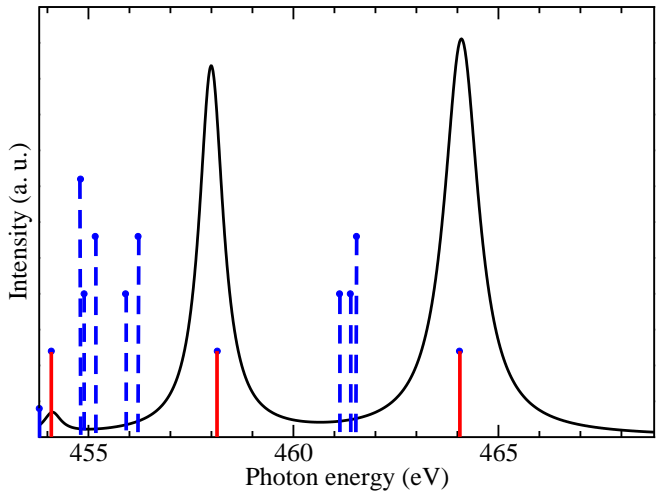}
\caption{(Color online)  
In red and blue $2p^53d^1$ multiplet energy levels. The length 
of the bars is proportional to the degeneracy (in red $J=1$). In black 
XAS L-edge spectrum for a Ti$^{4+}$ free-ion~: we see that only the 
final states with a $\Delta J=0,\pm 1$ with respect to the ground state 
contribute to the spectrum. 
}\label{Ledge}
\end{figure}

The incoming photon is defined by its polarization ${\bm \epsilon}$ and its 
wavevector ${\bf k}$. If we assume that the vector potential can be expanded 
in plane-waves, keeping 
in mind that ${\bf k}\cdot{\bf r}\ll 1$ we can write the 
matrix elements we need to compute for the transition like~:
\begin{equation}
\langle\psi_i|{\bf\epsilon}\cdot{\bf p}e^{i{\bf k}\cdot{\bf r}}|\psi_0\rangle
\sim
\langle\psi_i|{\bf\epsilon}\cdot{\bf p}
\left(1+i {\bf k}\cdot{\bf r} - ...\right)|\psi_0\rangle
\end{equation}
In the electric dipole approximation we replace the expansion by 1, such that 
the operator is given by~:
\begin{equation}
\hat{\mathcal O}= {\bf\epsilon}\cdot{\bf p}={\bf\epsilon}\cdot
[{\bf r},{\mathcal H}]
\propto {\bf\epsilon}\cdot{\bf r}
\end{equation}
Using an expansion on the $Y_1^m$, we finally obtain~:
\begin{equation}
\hat{\mathcal O}\propto r\left(\epsilon_1 Y_1^1+\epsilon_0 Y_1^0
+\epsilon_{-1} Y_1^{-1}\right)
\end{equation}
where the coefficients $\epsilon_i$ represent the projection of the 
polarization vectors on the $Y_1^m$ basis.  

At the dipolar approximation level, there are some selection rules 
for the transition which involve different quantum numbers~:
\begin{equation}
\Delta l=\pm 1\ ;\ \Delta s=0\ ; \Delta J=0,\pm 1 ;\ ...
\end{equation}
It should be mentioned here that these rules are automatically included 
in our code through Clebsch-Gordan and Gaunt coefficients when 
we compute the overlap. 

As an example, in Fig.\ref{Ledge} we take the standard case of an L-edge 
absorption for a Ti$^{4+}$ ion~: $2p^63d^0 \to 2p^53d^1$.
The effect of the dipolar selection rules is clearly seen~: 
since the ground-state $|\psi_0\rangle$ has a total momentum $J=0$, only 
the three states with $J=1$ in the multiplet of $|\psi_i\rangle$ contribute 
to the formation of peaks in the XAS spectrum. The inclusion of a 
crystal field would change the character of the states in the multiplet, 
allowing therefore for new dipole transitions and the appearance of new 
peaks.

\subsection{Comparison to experiments : ATiO$_3$ }
\begin{figure}[t]
\includegraphics*[width=\columnwidth,angle=0]{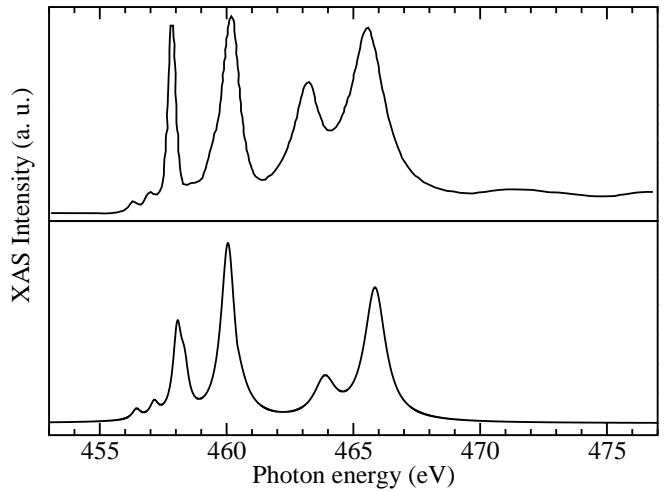}
\caption{XAS spectrum for a purely $2p^6 3d^0$-case: SrTiO$_3$ at 
the titanium L-edge. Upper panel: experimental result obtained by 
J. Schlappa {\it et al.} Ref.[\onlinecite{schlappa}]. 
Lower panel: multiplet 
calculation}\label{srtio3xas}
\end{figure}
\begin{figure}[t]
\includegraphics*[width=\columnwidth,angle=0]{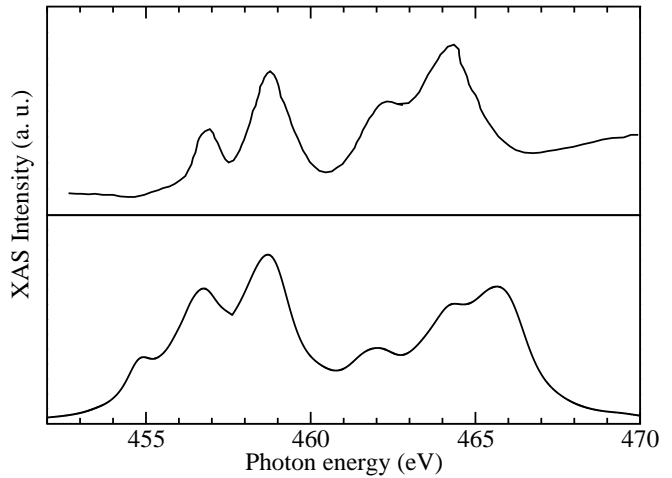}
\caption{XAS spectrum for a $2p^6 3d^1$-case: LaTiO$_3$ at 
the titanium L-edge. Upper panel: experimental result obtained by 
T. Higuchi {\it et al.} Ref.[\onlinecite{higuchi}]. 
Lower panel: multiplet 
calculation}\label{latio3xas}
\end{figure}

ATiO$_3$ (A=Sr,La) are perovskite structure materials: 
the Ti ions are in a cubic environment. Since Sr and La carry a formal charge 
$2+$ and $3+$ respectively, 
at the titanium L-edge the optical transitions 
are given by $2p^63d^0 \to 2p^53d^1$, and $2p^63d^1 \to 2p^53d^2$ 
respectively. The corresponding absorption spectra are displayed in  
Figs.\ref{srtio3xas} and \ref{latio3xas}

It is relatively straightforward to understand the $2p^63d^0 \to 2p^53d^1$ 
XAS spectrum of Fig.\ref{srtio3xas}: the two groups consisting of two large 
peaks correspond to the $L_3$--$L_2$ splitting while the peaks themselves 
represent $t_{2g}$ ($\sim 458eV$ ; $\sim 463.5eV$) and $e_{g}$ 
($\sim 460eV$ ; $\sim 465.5eV$) levels. The two small satellites close 
to $\sim 457 eV$ are due to the fact that the final states are in fact 
a bit more complicated than this simplistic view~: to the $t_{2g}$--$e_g$ 
splitting occurring for the $3d^1$ configuration we should also consider 
the effect of the $2p^5$ shell, which will give rise to these features. 

For the $2p^63d^1 \to 2p^53d^2$ transition we still can observe a 
$L_3$--$L_2$ splitting on Fig.\ref{latio3xas}, as well as a 
$t_{2g}$--$e_g$ splitting although it is less clear than in the previous case 
since here, the effect of electron-electron interaction tends to give 
rise to a more complex multiplet structure. 
Furthermore, LaTiO$_3$ is known to be a Mott-Hubbard insulator,\cite{imada} for 
which the electronic correlations play a more important role 
than in SrTiO$_3$, therefore the present fully ionic approach partly fails 
to reproduce the XAS spectrum in a truly quantitative way.

Although our present approach does not take charge fluctuations explicitly 
into account we still can mimic their impact on an absorption spectrum 
whenever the charge fluctuation is relatively weak. 
To this effect and as example, we consider in this paragraph 
a slightly doped case for these titanates 
compounds:  Sr$_{0.9}$La$_{0.1}$TiO$_3$. 
The wave-function for the ground-state will then consist in a linear 
superposition of the two previous one~: 
$|\Psi_0\rangle = 0.9 |3d^0\rangle + 0.1 |3d^1\rangle$. 
\begin{figure}[t]
\includegraphics*[width=\columnwidth,angle=0]{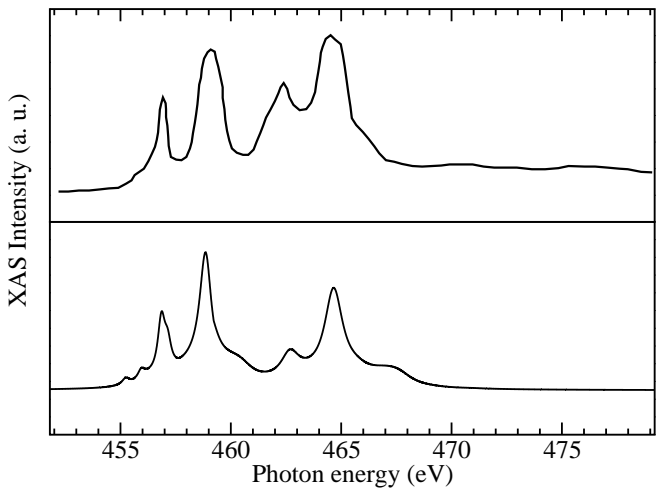}
\caption{XAS spectrum for a doped-case. 
Upper panel: experimental result obtained by 
T. Higuchi {\it et al.} Ref.[\onlinecite{higuchi2}]. 
Lower panel: multiplet calculation arising from the superposition
0.9$I_{d^0}$+0.1$I_{d^1}$.}\label{dopedtio3xas}
\end{figure}
The result is compared to the experimental data taken by T. Higuchi 
{\it et al.} in Ref.[\onlinecite{higuchi2}]. Although the writing 
of the wave-function as a superposition is formally correct, the 
evaluation of the XAS spectrum from a superposition remains an approximation, 
neglecting interference terms. 
The expression of Eq.(\ref{abs1}) would contain interference terms which 
vanish in the present approach. The approximation gives a base to  
describe the experiment, by comparing to the undoped case of 
Fig.\ref{srtio3xas}, we can see that the effect of the doping largely 
contributes to the formation of satellites peaks around the main features 
which were previously addressed to $e_g$ and $t_{2g}$ levels, this 
effect leads to a change in the relative weight and intensity of each peak. 
Thus a direct and systematic comparison of the computed XAS spectra for 
different doping levels can give insights on the respective role of the 
$e_g$ and $t_{2g}$ bands that are formed and contribute to experimental 
XAS data.  

\section{Resonant Inelastic X-Ray Scattering (RIXS)}
Resonant Inelastic X-ray Scattering (RIXS) is a 
photon-in photon-out spectroscopy with an incoming photon energy 
$\hbar\omega_{in}$ close to an absorption edge,  
such that the denominator of the second order-term in a Fermi golden-rule 
is small  and therefore completely dominant. Being of second order nature, 
we see immediately 
that the RIXS process involves not only an initial $|\psi_0\rangle$ 
and a final $|\psi_f\rangle$ state but 
also an intermediate state $|\psi_i\rangle$ with a finite 
core-hole lifetime $\Gamma_i$ which can be taken as a constant. 

In a first time, starting from the ground-state of energy $E_i$, 
the intermediate state, at energy $E_n$, is accessed {\it via} an optical 
excitation and hence depend on the polarization the incoming light 
$\epsilon_{in}$. 
The second part of the process consists 
in the de-excitation from the intermediate state to a final state of energy 
$E_f$, in emitting a photon of 
energy $\hbar\omega_f$. This process results in the writting of the 
Kramers-Heisenberg formula~:
\begin{equation}
I
\propto
\displaystyle\sum_f \left|
\displaystyle\sum_i \frac{\langle \psi_f|{\mathcal O}^\dagger|\psi_i\rangle
\langle \psi_i|{\mathcal O}|\psi_0\rangle}
{E_i-E_0-\hbar\omega_{in}-i\Gamma_i}\right|^2
\delta\left(E_f-E_0-\hbar\Omega\right) 
\end{equation}
where the same optical selection rules as in the previous paragraph 
will apply. In the later expression we see that the intensity also depends 
on the energy-loss $\hbar\Omega=\hbar (\omega_0-\omega_f)$. 
The experimental RIXS spectra also contain broad fluorescence lines
which mainly come from the emission from intermediate states with 
finite lifetimes which are not taken into account in the multiplet 
calculation. It should be mentioned that, apart from the fluorescence, 
the broadening of the $\delta$-function, in the energy loss direction, 
depends on the lifetime of the final state $|\psi_f\rangle$ and on the 
experimental resolution, whereas the broadening in 
the incoming photon energy direction is mostly governed by the lifetime of 
the core-hole intermediate state.

\begin{figure}[t]
\includegraphics*[width=\columnwidth,angle=0]{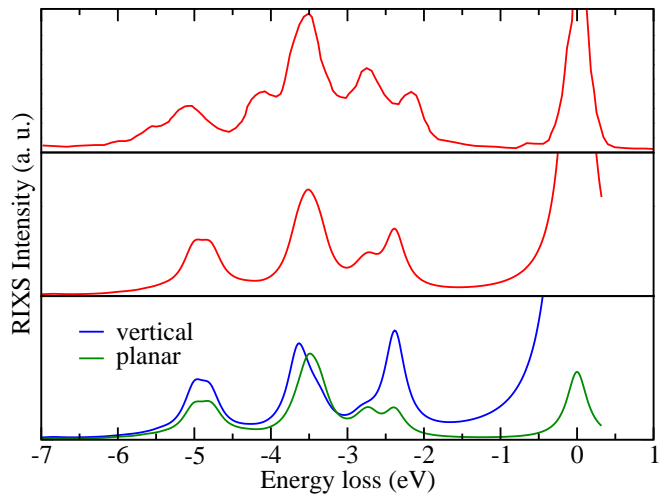}
\caption{(Color online) 
Upper panel: RIXS spectrum on MnO taken at the Mn L-edge, 
data by G. Ghiringhelli {\it et al.} Ref.[\onlinecite{ghiri}].
Middle panel: spectrum obtained with 
the multiplet approach for unpolarized outgoing photons.  
Lower panel: spectra obtained with the multiplet code. The different colors 
are assigned to different polarizations of the outgoing photon: 
in-plane and in vertical polarizations.}\label{mnorixs}
\end{figure}

In order to have a specific example we compare with the Mn L-edge data 
by Ghiringhelli {\it et al.} for MnO.\cite{ghiri} 
From a structural point of view, 
MnO has the same crystal structure as NaCl, on the charge distribution side, 
although MnO is not purely ionic, our code 
describes a static Mn$^{2+}$O$^{2-}$ picture, however we can adapt and scale 
the crystal field by changing the formal charge of the ligands. Hence, at the 
L-edge, we have transitions of the type: 
$2p^6 3d^5 \to 2p^5 3d^6 \to (2p^6 3d^5)^\star$.

In Ghiringhelli {\it et al.}'s paper,\cite{ghiri} 
the data were obtained for two different polarizations of the incoming photon: 
in the scattering plane and in the direction normal to the plane. 
We take the same geometry for our simulation with a plain MnO$_6$ octahedron. 
To first compute the multiplet, we consider an Mn ion with a formal neutral 
charge to avoid any shrinking of the wave-function 
and 6 O$^{2-}$ ions with a formal charge $2-$, the Mn--O 
distances are taken to be $\sim 2.2$ \AA$\ $ and the octahedron is assumed 
to be undistorted. The results, scaled by $80\%$ and with a constant 
broadening of 0.2 eV in the energy-loss direction, 
are shown in Fig.\ref{mnorixs} and compared to the actual experimental data. 
As shown in the lower panel of the figure, 
we have the possibility to investigate the polarization dependence for the 
outgoing photon this offers the opportunity to study the character of the 
different excitations depending on how each peak is affected by the 
selection rules. For instance, the peak close to an energy-loss of $\sim 4$eV 
is more proeminent for crossed polarizations. 
The effect of the polarization is also clearly visible on the elastic peak 
which weakens considerably~: since the current effort in the study of 
transition metal oxides tends to investigate low-energy physics 
it would be of interest to take advantage of this effect to study the physics 
occurring in the vicinity of the elastic peak.

\section{Conclusion and Outlook}
In summary, we have described an easily accessible approach for the
computation of multiplet spectra arising in for soft X-ray 
spectroscopies of narrow band solids. 
The principal input specification is for the element where the core is to be excited
and the core and valence configurations involved in the transition 
(for example: $2p^6$ $3d^2$ ). 
The central field part of the present theory is done with a Dirac relativistic implementation
for a simple density functional, and yields thus the spin-orbit splitting from first principles.
The electron-electron interaction amongst the open shell orbitals is obtained in
the subspace defined by the configuration specification.
An integral part of the approach is the treatment of arbitrary 
crystal fields. The specification of the crystal field
is via straightforward specification of coordinates and charges for a small number of 
neighbor ions. The crystal field input remains easily overseeable for 
arbitrary symmetry cases, without requiring expert knowledge in group notation.  
Further fundamental inputs are the core hole lifetime, and  polarization of the light
with respect to the coordinate system used for the crystal field.
With this minimal input, the method constitutes a first principles approach, albeit with approximations.
Comparison with experimental multiplet spectra is quite good usually without any fitting
efforts.

Nevertheless, the method can be tuned by a number of empirically adjustable parameters
to fit experiment more accurately. The rationale for the empirical parameters are
screening effects and coupling to energy bands that are not in the scope of the model.

In its first version the method allows to calculate X-ray absorption spectra (XAS)
and Resonant Inelastic X-ray Scattering spectra (RIXS). 
Comparison with experimental spectra was presented in this article for some test cases.
Planned extensions should include non ground state
configurations in the RIXS. Extension to multiplets in photo electron spectroscopy
and circular magnetic dichroism spectra is considered.
Inclusion of charge transfer effects adjusted {\it via} 
empirical input parameters may also be considered in the near future.

We plan to make this type of calculations widely accessible in the near future 
{\it via} a web interface.

\vskip2.cm

\acknowledgements
We deeply acknowledge our colleagues from the ADRESS beamline at the Swiss 
Light Source: Thorsten Schmitt, Justina Schlappa, Kejin Zhou and 
Luc Patthey for their constant support and valuable inputs during our 
discussions. 

We gratefully acknowledge D. D. Koelling for the Dirac central field 
radial equation solver. 

We also thank C. Dallera, C. McGuiness and 
J. Nordgren for stimulating discussions and 
encouragement.

\appendix
\section{Double counting}
The electron-electron interaction is given 
by the standard Coulomb repulsion, with the following expansion:
\begin{equation}\label{electro}
\begin{array}{rcl}
V_{e-e} &=&\frac{1}{|\bm{r_1 - r_2}|} \\
\ \\
&=& \displaystyle\sum_{l=0}^\infty
\displaystyle\sum_{m=-l}^{l}\frac{4\pi}{2l+1}
\frac{r_1^l}{r_2^{l+1}}
Y_l^m(\theta_1,\phi_1)^\ast Y_l^m(\theta_2,\phi_2)\ \\ 
\ \\
&{\rm with}&\ \ r_1<r_2
\end{array}
\end{equation}

Two terms appear in equation (\ref{hamilton})~: electrostatic interaction 
as well as eigenenergies coming form the solution of Dirac equation 
$\epsilon_i$. Through the self-consistentdetermination of the effective 
potential $V_{eff}$ appearing in the Dirac equation, 
these last terms contain already part of the electrostatic 
interactions: we thus should be careful and avoid double-counting.  
This can be achieved in subtracting the already counted part to 
Eq.(\ref{electro}) by changing the sum 
$\sum_{l=0}^\infty \to \sum_{l=1}^\infty$. The $l=0$ term can be formally 
viewed as a $\sim 1/r$ -term which is already treated in the potential 
of the Dirac equation. 

It is also worth mentioning that whenever a shell is full one can simply 
drop the electrostatic terms involving that shell since it only results in a 
common shift of the eigenenergies.


\end{document}